\documentstyle[11pt,aaspp,epsfig,psfig]{article}
\begin{document}

\title{Distribution of Spectral Characteristics and the Cosmological
Evolution of GRBs.}
\author{Nicole M. Lloyd and Vah\'e Petrosian}
\affil{Center for Space Sciences and Astrophysics, Stanford University,Stanford,
CA 94305-4060}

\begin{abstract}

 We investigate the cosmological evolution of GRBs, using
 the total gamma-ray fluence as a measure of the burst strength.
 This involves an understanding of the distributions of the spectral 
 parameters of GRBs as well as the total fluence distribution - both
 of which are subject to detector selection effects.  We present new
 non-parametric statistical techniques to account for these
 effects, and use these methods to estimate the true
 distribution of the peak
 of the $\nu F_{\nu}$ spectrum - $E_{p}$ - from the raw
 distribution. The distributions are obtained 
 from four channel data and therefore are rough estimates; hence, we
 emphasize the methods and present qualitative results.
 Given its spectral parameters,
  we then calculate the total fluence for each burst, and compute its cumulative
 and differential distributions.  We use these distributions to estimate
 the cosmological rate evolution of GRBs, for three cosmological
 models.  Our two main conclusions are the
 following:  1) Given our estimates of the spectral parameters,
we find that
there may exist a significant number of high
$E_{p}$ bursts that are not detected by BATSE, 2)
We find a GRB co-moving rate density quite different from that
 of other extragalactic objects; in particular, it is different
 from the recently determined star formation rate.

 {\it Subject headings}: gamma rays: bursts -- methods: statistical --
 cosmology: observations -- miscellaneous.
\end{abstract}

\section{Introduction}
   Identification of several gamma-ray bursts (hereafter GRBs) with X-ray,
   optical and radio afterglows, as well as measurements of the redshifts of
   GRB 970508, GRB 981214, and GRB 980703 are beginning to
   define the GRB paradigm.  However, this information is not sufficient
   for detailed cosmological studies of their distribution and evolution.
  Until more information on the distances to GRBs are
revealed, we must continue to rely on standard statistical
studies (such as obtaining number count distributions of fluence
or flux)
to gain further insight on the nature of these events.  
    
      However, information on the spatial distribution - although necessary -
will not be sufficient for solving the puzzle of GRBs.
In particular, we need spectral information to 
understand the energy release and
radiation processes at the burst.
This paper deals with the determination of the spatial and spectral
properties of GRBs.

 The information on the spatial distribution
comes primarily from the investigation of source counts or the log$N$-log$S$
relation.  Assuming a luminosity function (LF) and a cosmological
model, one can convert such
counts to spatial (or redshift) distributions (Weinberg, 1972).  The likelihood of
the assumed luminosity function determines our confidence in the redshift
distribution or cosmological evolution of these sources.
The usual practice is to use the peak flux $f_{p} (\nu)$ as a measure of
the burst strength.  This flux
is related to the peak luminosity $L_{p}(\nu)$ via $L_{p}(\nu) = f_{p}(\nu)
 \Omega d_{L}^{2}K(z)$, where $d_{L}(z)$ is the luminosity distance, $K(z)$
 is the so-called K-correction due to redshift of the spectrum, and $\Omega$
 is the solid angle into which the radiation is beamed ($\Omega = 4 \pi$ for isotropic
 emission). 
  However, we have no knowledge of the burst
peak luminosity function - say $\Psi (L_{p}(\nu ))$ - 
and it is unclear what assumptions
can be made about this function.  For example, in a fireball model,
 both the duration
and the peak luminosities are expected to depend on the details
of the fireball (e.g. the bulk Lorentz factor and amount of baryon
loading), so that their distributions will be broad and perhaps
comparable to the observed range of the peak fluxes. As a result,
the luminosity function cannot be readily de-convolved from the spatial
distribution, and we cannot get an accurate measure of the burst
evolution or spatial distribution from the log$N$-log$S$ diagram. 
  
   On the other hand, in a neutron star (or black hole)
merger model, the total energy released - represented by the
gravitational potential energy of the merging process - is
expected to be well defined and have a narrow distribution; this is
also most likely true for the ``hypernova'' model.  Hence, we would like
to determine 
what radiative signature is 
a representative measure of this energy and the strength of
the GRB.

As the name ``gamma ray burst'' implies, 
the peak gamma-ray flux $f_{p}$ (or luminsity $L_{p}$) of a GRB
far exceeds the
peak fluxes at other energies.  However, until recently, we did
not know at what frequency the total
radiated energy peaked.
The recent X-ray, optical and radio observations 
of the afterglows - in particular the fact that these fluxes decline
more rapidly than $1/t$ (see e.g. Murakami et al., 1997, Galama et al., 1997)
indicate that the energy {\bf fluence} in the gamma-ray
range is also higher than in any other band.  For isotropic
emission or a beaming angle independent of wavelength, this implies
that the total gamma-ray radiated energy, $\cal E$$_{tot}$, is the best measure of
the total released energy. Therefore, it is plausible that  
the distribution of the gamma-ray radiated energy is narrow, so that 
the total gamma-ray
fluence $F_{tot} = \cal E$$_{tot}(1+z)/(\Omega d_{L}^{2})$ 
provides the best indication of the strength of the burst for
the log$N$-log$S$ analysis and other investigations (assuming $\Omega$
has a narrow distribution).
In particular, it is a better measure
than the observed fluence $F_{obs}$, which represents the fluence
in some narrow energy range (i.e. the energy
range in which the detector triggers).
  
Note that the relation between the
monochromatic flux
(or fluence) and the corresponding luminosities require a knowledge of the 
so called K-correction due to redshift of the spectrum. This complication
is not present in the relation between the total radiated energy and
the total gamma ray fluence.
  It is also important to note that $F_{tot}$
depends on the
spectral properties of the burst.  Hence, if we want to simplify matters
 by dealing with
the total fluence, we need to understand the spectral distribution
$F(\nu)$ of the GRBs.  

  The primary goal of this paper is to demonstrate how we
  can obtain information on the cosmological
evolution of GRBs, using the observed distribution
of $F_{tot}$.  However, for an accurate determination
of the distribution of $F_{tot}$, we need to determine
the spectral parameters of each burst.  It is common to use 
the following four parameters (say, in a Band spectrum (1993)) to
characterize a burst's spectrum:
 the photon energy at the peak of the $\nu F_{\nu}$
spectrum, $E_{p}$,  the low energy spectral index, $\alpha$,
the high energy spectral index, $\beta$, and a normalization $A$.
  The best source of
data for the purpose of determining these parameters
and $F_{tot}$ is the BATSE instrument on board the Compton
Gamma-Ray Observatory (CGRO).  However, because the BATSE
trigger criterion is based
on photon counts between 50 - 300 keV (for the 3B catalog), the
selection
biases or thresholds on $F_{tot}$ as well as the spectral parameters
can be quite complicated.
In \S 2 we show how the trigger criterion used by
BATSE can be used to obtain the selection bias on any measured
quantity - in particular, $F_{tot}$ and $E_{p}$ - and
describe new methods to account for the resultant complicated
data truncation.  In \S 3, we present our results on the determination
of the spectral parameters. The distribution
of $E_{p}$ is given in
 \S 4, and in \S 5 we derive the distribution of $F_{tot}$, give a scenario
for cosmological evolutions of the GRBs, and discuss its relation to the 
new observations of afterglows and evolution of other cosmological
sources.  A brief summary is presented in \S 6.

\section{Description of Data and The Basic Problem}
\subsection{The Data}
   BATSE has produced a large catalog of GRBs with a wealth of 
   information on their temporal and spectral characteristics.  This
   detector has been designed to optimally detect GRBs and has 
   been very successful in achieving this goal.  Nevertheless, like
   all detectors, it has intrinsic limitations and may miss certain
   types of bursts or provide only limited information on others.  BATSE
   is a triggered instrument; it will record data on a burst if
   the detector counts in a time interval $\Delta t$ and energy
   range $\Delta E$ exceed some threshold value $C_{min}(\Delta t)$
   determined by the background fluctuations.  BATSE uses three
   trigger time intervals $\Delta t = 1024, 256,$ and $ 64$ ms with
   four energy channels $\Delta E = 25-50,50-100,100-300,$ and $ > 300 $ keV.
   For most of the data, triggering is based on counts in channels
   2 and 3; recently, however, other channel
   combinations have been tried.  Here we will use the
4 channel LAD fluence values published in the BATSE 3B catalog (Meegan
et al., 1995), which utilizes the former trigger scheme, so that the observed fluence
$F_{obs} \equiv \int_{50keV}^{300keV} F(\nu ) d\nu$.  However, the formalism
described below is applicable to the latter data as well.
Note that we analyze bursts with $C_{max}/C_{min}$ on 1024ms 
timescale.  Because of the lack of resolution of the four channel
data, here we will emphasize the method of our calculations
rather than give concrete quantitative results.
Future work
will involve calculations made from 16 channel CONT data.

\subsection{Data Truncation}  
      The BATSE detector trigger conditions lead to several limitations
      on the data.  The most obvious is that only bursts with peak counts 
     $C_{max} > C_{min}$ are detected.  Determination of the average 
 $V/V_{max} = (C_{min}/C_{max})^{3/2}$, and its distribution were one of the
 primary objectives of BATSE using this trigger configuration (Meegan
 et al., 1992).     
  Another limitation
     arises from the finite durations $\Delta t$ of the triggering
     criteria.  As pointed out by Petrosian and Lee (1994, see also
     Lee and Petrosian, 1996), this reduces
     the detector efficiency of short duration bursts (duration $T < \Delta t$)
     in the sense that there is a relative bias against detection of weaker
     bursts with shorter duration.  Similarly, as pointed out by Piran and
     Narayan (1996), the finite 
 range $\Delta E$ of the detector energy channels
     provides another limitation (see also Cohen, Piran, and
     Narayan, 1998).  In this case, the detector efficiency
is highest for bursts with most of their photons in the range $\Delta E$, 
and decreases for bursts with more of their flux outside this range.

 It turns out that we can account for
these limitations by utilizing the BATSE trigger condition:
 Let us consider a burst characteristic, say X,
 whose value is measured by BATSE.  For each burst, then, we
 know X, $C_{max}$, and $C_{min}$.
 Now, in the spirit
of the $V/V_{max}$ test, one can ask what is the possible range of 
values of X that this burst can have and still trigger the instrument. 
In general, X can have
an upper limit $u$, a lower limit $l$, or both limits; $X  \in T =
[l,u]$.  This question was first
posed by Petrosian and Lee (1996) in connection
with the determination of the detection threshold or efficiency of
the observed energy fluence $F_{obs}$; in this case,
the fluence has only a lower limit $F_{obs,lim}$.
 It was shown that this threshold
(or for that matter, the
threshold on any other measure of burst
strength which is proportional to the photon counts)
is obtained from the simple relation 
\begin{equation}\label{equation}
F_{obs}/F_{obs,lim} = C_{max}/C_{min},
\end{equation}
provided the burst's spectrum does not change drastically
throughout its duration.  A similiar relation holds for the
total fluence $F_{tot}$ as we describe in \S 5.

For other bursts characteristics, this may not be the case.
For example, we have shown (Lloyd and Petrosian 1998) that the spectral
parameter $E_{p}$ has both an upper and lower limit.  The values
of $E_{p, max}$ and $E_{p, min}$ are not related to $C_{max}/C_{min}$
and $E_{p}$ in a simple form as above, and thus we require a
more complex procedure to determine
their values.  We will describe this in \S 3. Given the thresholds,
we can then use non-parametric statistical techniques 
 to obtain an estimate the
 parent distribution of the relevant observable from the
observed distribution.

\subsection{Dealing with Data Truncation}
  Correcting for data truncation in the one-sided case was
  discussed in detail by Efron and Petrosian (1992).  This method
  has been used to determine the correlation between 
  parameters in a bivariate (or multivariate) setting, and
  to obtain univariate distributions of several
  GRB characteristics (see, e.g., Lee and Petrosian, 1996).
  The basic idea is that one can use information from the
  untruncated region of a distribution to estimate how much
  data is missed due to the truncation.  For a description
  of the basic process, see also Petrosian (1992 or 1993).
  
  Dealing with data which has both upper and lower limits is more
  complex and requires
  a generalization of the above technique.
    Recently, Efron and Petrosian (1998) have developed a
  method
to deal with this kind of data  truncation.  We briefly review
this method here.  

Consider data points $(x_{i},y_{i})$ (in our case $(E_{p},F_{obs}))$,
where $x_{i}$
has lower and upper limits  $l_{i}$ and $u_{i}$, respectively;  
$x_{i}  \in T_{i} = [l_{i},u_{i}]$.  Let us also assume that
$x_{i}$ and $y_{i}$ are uncorrelated.
Let $g(x)$ be the true distribution of $x$, which
would be observed if there were no truncations. 
However, because $x_{i}$ is limited to $T_{i}$, we observe the
conditional distribution $g(x_{i}|T_{i}) = g(x_{i})/G(T_{i})$, where 
$G(T_{i}) = [\sum_{j} g(x_{j})$ : $x_{j} \in T_{i}$] 
is the probabilitly that $x$ exists in
$T_{i}$. We normalize $g(x_{i})$ such that $\sum_{i=1}^{N} g_{i} = 1$,
and define the vectors

\begin{equation}\label{equation}
g_{i} = g(x_{i}) \ \  {\rm and }
\ \  G_{i} = G(T_{i})
\end{equation} 
and the matrix
\begin{equation}\label{equation} J_{i j} = \cases{1,&  $x_{j} \in T_{i},$\cr
0,&  $x_{j} \notin T_{i}.$}
\end{equation} 
Furthermore, by definition, one can show that 
$G_{i} = J_{i j} g_{j}$ (where
we have used the convention that repeated indices are summed over).
The goal is to estimate $g(x)$ from $l_{i},u_{i}$, and $x_{i}$
assuming all $N$ cases are
independently distributed.  It can be shown that maximizing the likelihood
(see Maloney et al., 1998) of the observed data set gives
\begin{equation}\label{equation}
{1/g_{i} = J_{j i}*(1/G_{j}) + \rm const, }
\end{equation}
where $J_{j i}$ is the transpose of $J_{i j}$.
The procedure amounts to solving this equation
iteratively, under the normalization condition and definition
of $G_{j}$ and $J_{i j}$ given above.
Convergence is reached when $const$ in the above equation goes to zero.
This method can be used to determine univariate cumulative ($G$)
 and differential ($g$)
distributions.  As described in Efron and Petrosian (1998),
it can also be used to determine correlations between
relevant variables (e.g. fluence and $E_{p}$ or peak flux and $E_{p}$).

  We will use the methods described above to correct
for data truncations and produce
bias free distributions of the peak energy
 $E_{p}$ and the total fluence $F_{tot}$; we will then directly
 use the latter distribution to investigate the cosmological
 evolution of GRBs.

\section{Spectral Characteristics}

The spectral properties of GRBs provide the most direct information about the 
physical processes associated with the event. In particular, 
the distribution of the spectral
parameters and the correlation between these and 
other GRB characteristics can shed 
significant light on the radiation 
mechanisms and energy production of the burst. 
 
Therefore, our first task is to determine
 the spectral parameters of each burst.
 We characterize the  
  spectral forms of the bursts by four parameters - 
$\alpha$ and
$\beta$ (the photon power law index at low and high
energies respectively), $E_{o}$ (a break energy related
to the peak of the $\nu F_{\nu}$ spectrum)
 and $A$ (a normalization in units of 
$ergs/cm^{2}$).  We choose two functional forms to represent the
{\bf energy} fluence spectra
- a Band spectrum, and a generic double
power-law spectrum with a smooth transition between the low energy and high
energy power law behavior: 

  \begin{equation}\label{equation}    F(E) =
\cases{ (A/E_{o}) (E/E_{o})^{\alpha+1} {\rm exp}[(\beta-\alpha)(E/E_{o}-1)],&  
  $ E < E_{o}$ \cr
(A/E_{o}) (E/E_{o})^{\beta+1},& 
$E >  E_{o}$}
\end{equation}

 \begin{equation}\label{equation}    F(E) =
\frac{(A/E_{o})(E/E_{o}))^{\alpha+1}}{1+(E/E_{o})^{\alpha-\beta}}
\end{equation}
We are interested in bursts whose $\nu F_{\nu}$ spectra have a maximum,
so that we may correctly speak
of $E_{p}$ as the  ``peak of $\nu F_{\nu}$ ''.  To
define $E_{p}$ (as well as to keep the burst's total fluence $F_{tot}$
finite), we require $\alpha$ to be greater than -2 and $\beta$ to
be less than -2 (which also implies $\alpha > \beta$).  As
we discuss below, this does not impose
any large bias in our sample. Then, with these
constraints, we can define $E_{p}$ for each
spectrum above:
 $E_{p} = E_{o}(\frac{\alpha +2}{\alpha-\beta})$
 and 
$E_{p} = E_{o} (\frac{-(\beta+2)}{\alpha+2})^{\frac{1}{\beta - \alpha}}$
for (5) and (6), respectively.  

 From four channel data, we can determine the
  four spectral parameters with no degrees of freedom.  However,
  this introduces strong interplay between $E_{p}$ and $\beta$ or
  $\alpha$.  We avoid this by selecting $\beta$ from
an assumed gaussian distribution with a mean of $\approx -3$, and
a standard deviation of $1$ (and of course the constraint $\beta < -2$).
 This
allows for one degree of freedom. In fact, we have performed this analysis
using several distributions for $\beta$ (i.e. a uniform distribution,
or a gaussian with a different mean and standard deviation), and have
found similiar qualitative results. However, we should note that
steeper, more negative $\beta$'s will naturally produce higher $E_{p}$'s,
to accommodate the fluence in high energy channels (3 and 4).
We
find the rest of the parameters by minimizing $\chi ^{2}$ via
a downhill simplex routine (e.g. see Press et al., 1992).
For the Band spectrum, we find acceptable fits for 291 bursts, while
268 give acceptable fits for the second spectrum listed above,
out of a total of 518 bursts.
Figures 1(a) and 1(b) show
 the raw distributions of the spectral parameters obtained
with this method for the two spectral forms in equation 5 
and 6, respectively. 
 
 It turns out that the distribution of the parameters for bursts
   with unacceptably high $\chi^{2}$ is not very different
   from those with acceptable $\chi^{2}$.  We have tested
   to see if the reason for many high $\chi^{2}$ bursts
   came from the constraints on $\alpha$ and $\beta$ mentioned
   above.   Relaxing these restrictions,
 we found that only $\sim 1 \%$ of bursts that gave
 acceptable fits had an $\alpha < -2$, $\beta > -2$, or $\alpha > \beta$.
 We therefore assume that the bursts with
  good fits are representative of the total sample.
 However,  because the 
lack of energy resolution of the four channel
data, we would like to emphasize that our fits are rough
estimates of the actual spectral parameters, so that the following 
analysis exhibits our method rather than presenting a final
quantitative result. 
As mentioned previously, in future work we will use fits from
16 channel CONT data, which will provide a more accurate determination
of these parameters.

Several studies have obtained the spectral parameters for the brightest GRBs 
(e.g., Band et al. 1993) and investigated correlations between these
parameters and peak flux,
duration and
spatial distributions (Mallozzi et al. 1995, Kouvelioutou et al. 1995,
Pendleton et al. 1996).
However, caution is required in such studies, when
dealing with the raw data. The effects of selection biases must be
evaluated before
 obtaining these distributions and correlations. 
 The parent distributions of the spectral
 parameters could be quite different
from the observed distributions shown in the
above figures.  The differences are caused by the truncations
resulting from the 
triggering procedure, and are a measure of trigger efficiencies for
each of the four spectral parameters $\alpha, \beta, E_{p}$, and $A$
or $F_{tot} = A \xi (\alpha, \beta)$.
The truncation on $A$ or $F_{tot}$ is described
by equation (1): $F_{tot,lim} = 
F_{tot}(C_{min}/C_{max})$.  As 
mentioned previously and shown in detail in the
next section, there are also selection biases against high and low values of 
$E_{p}$. For instance, 
BATSE is most likely to see bursts with $E_{p}$ 
in the energy range in which the detector triggers. 
Becuase the detector needs some minimum
number of counts $C_{lim}$ 
to trigger, then for each
burst one can find an interval $[E_{p_{min}},E_{p_{max}}]$ to which
$E_{p}$ is confined in order for the burst to have been detected.  
 
 Similarly, truncation is present for $\alpha$ and $\beta$.
For example, if a burst with given
$E_{p}$ and $A$ had a value of $\alpha$ which
was postitive and large or had a value of $\beta$ which was negative
and large,  - i.e. if the $\nu F_{\nu}$ spectrum had a steep rise
or fall - then the observed $C_{max}$ could
 fall below the
threshold $C_{min}$ and render the burst undetectable.  Using the methods
described in \S 2 and below, we can correct the raw distribution
of all four parameters and determine their correlations
with other GRB characteristics.  However, the effect of the truncation on 
$\alpha$ and $\beta$ is more complicated and not as pronounced as that
of $E_{p}$ and $F_{tot}$.  We will deal with the former in future publications.
Here, we limit our discussion to obtaining the true distributions
of $E_{p}$ and $F_{tot}$.
We will present the results from the Band spectrum fits
to the bursts only; the smooth double
power-law spectrum in equation (6) gives qualitatively
similiar results for all of the following analysis.

\section{Distribution of $E_{p}$}

 \rm For each burst we have the values
 of  $\alpha,\beta$, $A$, $E_{p}$,  and the observed
 fluence $F_{obs} = \int_{E_{1}}^{E_{2}} F_{\alpha,\beta,A}(E,E_{p}) dE$,
 with $E_{1} = 50 keV$ and $E_{2} = 300 keV$. 
 From equation 1, we can calculate the fluence
 threshold $F_{obs,lim}$. Using this limit, we can determine
 the possible range of values $E_{p}$ can take on, and still
 trigger the BATSE instrument. It is clear that the
 limiting value $E_{p,lim}$ must satisfy the equation
\begin{equation}\label{equation}
F_{obs,lim} = \int_{E_{1}}^{E_{2}} F_{\alpha,\beta,A}(E,E_{p,lim}) dE.
\end{equation}
For the spectral forms and parameter ranges discussed in
the previous section, the right hand side of equation (7)
increases monotonically with $E_{p,lim}$, reaches a maximum
for $E_{p,lim}$ within the limits of the integration, and then
decreases monotonically.  Consequently, there
will be two values of $E_{p,lim}$ that satisfy
this equation.  To find these two solutions, we start with the 
value of $E_{p}$ determined by the simplex routine (for which
the right side is equal to $F_{obs}$),
and increase it (while keeping 
 $\alpha$, $\beta$, and $A$ equal to their determined values)
until equation (7) is satisfied  - that is, until the observed
fluence is brought below the threshold.  This value
is the upper limit to $E_{p}$, $E_{p,max}$.
We then decrease $E_{p}$
until this equation is again satisfied.  This gives the lower
limit $E_{p,min}$.
In essence, the procedure
amounts to redshifting/blueshifting the burst until it is undetectable
by BATSE.  
Figure 2 shows $E_{p}$ vs. $E_{p,max}$ and
$E_{p,min}$ for the Band spectrum.  Notice the
truncation is much more prominent on the upper than the
lower end; hence, we expect to see a more significant correction to the high
end of the $E_{p}$ distribution.

Using the method described in \S 2.3 for doubly
truncated data, we corrected
the observed distribution given that the observations can detect only
bursts with $E_{p}$ limited to the interval
$[E_{p_{min}},E_{p_{max}}]$.    Figure 3
 shows the raw and corrected differential
distribution of $E_{p}$ for the Band spectrum.
The raw and corrected
distributions are normalized such that they 
are equal at low values of $E_{p}$ (where
the data truncation is inconsequential).  This normalization
visually emphasizes the differences of the distributions at high $E_{p}$.
The figures suggest that
a large number of high $E_{p}$
bursts are missed by the triggering procedure.   
  
In order to determine the significance of
the difference between the raw and corrected distributions, we carry out the
following two checks:  First, we perform a
K-S test on the cumulative distributions of $E_{p}$, and
a $\chi^{2}$ test on their differential distributions for both
the Band spectrum and the smooth double power law. For
the $\chi^{2}$ test, we divide the distributions into
two bins, with equal numbers of observed bursts per bin.
As shown in the table below, in both cases the tests indicate the
probability that the observed and corrected $E_{p}$ distributions
are the same is extremely small. 

\begin{center}
\begin{tabular} {lccr} \hline \hline
Test & Spectrum    &    Probability the distributions same \\ \hline
K-S   & Band       &    $1.8\times 10^{-6}$  \\
$\chi ^{2}$ & Band  &     $3.0\times 10^{-7}$ \\
K-S   & Double Power Law       &    $4.4\times 10^{-5}$  \\
$\chi ^{2}$ & Double Power Law  &     $2.4\times 10^{-5}$ 
 \\ \hline \hline
\end{tabular}
\end{center}

Second, we have carried out simulations to determine the
accuracy of the method described
in \S 2.3; Appendix A contains these results.
  We begin with a parent distribution
 of $E_{p}$'s, simulate an observational truncation (defined by
 the limiting fluence of the burst; see \S 4).  From this we obtain our
 observed distribution of $E_{p}$'s.  We then apply the above technique
 to correct for the truncation, and retrieve our parent
 distribution to very high accuracy, demonstrating the robustness
 of the method (see Appendix A
 for more details).
 
 Our results are
in qualitative agreement with the observational
data from the Solar Maximum Mission (SMM) presented by Harris and Share (1998a, 1998b);
SMM is sensitive to higher energies than BATSE, so that it
suffers the bias against detection of bursts
with hard spectra to a much lesser
degree.  The
results presented in Harris and Share (1998b) agree
qualitatively with our corrected $E_{p}$ distribution.  We perform a
similiar $\chi^{2}$ as above, comparing
the $E_{p}$ distribution from SMM with both the uncorrected
and corrected BATSE $E_{p}$ distributions presented in Figure 3.
In this case, we divide the data 
into two bins below and above 550 keV.  We find 
the probability that the $E_{p}$ distribution from SMM data
is the same as the raw BATSE $E_{p}$ distribution to be $1.3 \times 10^{-6}$.
On the other hand, we find a 90\% probability that
the $E_{p}$ distribution from the SMM data is the same as the
corrected BATSE $E_{p}$ distribution.  
Note that the statistical method we use
can correct the $E_{p}$ distribution only for the values
of $E_{p}$ observed by BATSE.  Therefore, we cannot determine
the distribution above $\sim 1$ MeV (where the raw BATSE
distribution ends).  Similarly, because
we have no observed $E_{p}$'s below $\sim 20$ keV, we cannot confirm
Strohmeyer et al. (1998)  reporting a significant number of
low $E_{p}$ bursts ($< 20$ keV) observed by GINGA, but not
observed by BATSE.  
 
 
 A bias against detection of high $E_{p}$ bursts has significant implications
on correlations of $E_{p}$ with other measures
of burst strength like peak flux, $F_{obs}$, or $F_{tot}$.
The detector is most likely
to miss the high $E_{p}$ bursts with low strength,
so that a simple correlation analysis
between raw values of $E_{p}$ and burst
strength without the consideration of the data truncation
can lead to erroneous correlations.
Examples of studies which may be affected by such a bias
include the correlation studies of 
Nemiroff et al. (1994) and Mallozzi
et al. (1995).  This will also have an important effect on
determining whether or not the correlations are intrinsic
or due to cosmological redshift (Brainerd, 1997).
  The
above results indicate that a more thorough analysis of the
correlations is required before conclusions can be reached
on the redshifts or intrinsic effects.  We will address this question in 
a future publication, when we have more accurate estimates
of the spectral parameters.

\section{Distribution of the Total Gamma Ray Fluence}
As discussed in the introduction,
 an important way to learn about the spatial distribution
of GRBs is to study the distribution of some standard candle
variable of the event.  We also conjecture the total radiated energy ${\cal E}$$_{tot}$
in the gamma-ray range
may be the best candidate for such a standard candle variable, and
that the total fluence $F_{tot}$ of the GRB may provide the best measure
of distance.
Consequently, we study the distribution of the bursts' total fluence; if
the burst radiates isotropically, the total fluence  
is related to the total radiated energy $\cal E$$_{tot}$ as:
\begin{equation}\label{equation}
F_{tot} = \frac{{\cal E}_{tot}}{4 \pi d_{L}^{2}}(1+z),
\end{equation}
where $d_{L}$, the luminosity distance, depends on redshift and
the cosmological model (Weinberg, 1972).

    Again, because the detector is sensitive only over
a finite energy range, BATSE does not
 necessarily obtain all photons from the burst.
   Hence, the detector measures only a portion of 
the burst's total fluence.  
However, the total fluence of a GRB can be obtained from the
spectral fits to the observed fluence simply by
integrating over the spectrum 
of that burst: $ F_{tot} = \int_{E_{min}}^{\infty} 
F_{\alpha,\beta,A}(E,E_{p}) dE.$, where $E_{min}$ denotes the
beginning of the gamma ray energy range (as an estimate, we
use $E_{min} = 25$ keV).
This fluence will have an observational lower limit in the same
way that the fluence from $50-300$ keV has, and  
 its threshold is given by
 the generalization
of equation (1): $F_{tot,lim} = \frac{C_{min}}{C_{max}} F_{tot}$.
Following the steps described 
in Petrosian and Lee (1996),  we first test for
any 
correlation between $F_{tot}$ and $F_{tot,lim}$  and parametrically
remove this correlation.  
We then use the method
described in Efron and Petrosian (1994) for
one sided truncated data (this is equivalent to the method
of \S 2.2, taking the upper limit $u$ to $\infty$), and   
obtain the cumulative and differential
distributions for the total fluence.

 Figures 4(a) and 4(b) show the cumulative and differential distributions
of the total and observed fluence for our sample.  Marked
on each curve are GRB 970508 (circle), GRB 971214 (cross), and
GRB 980703 (triangle).  Note that we do not find a good spectral
fit for GRB 980703; hence, we obtain a lower limit to its total
fluence simply by summing up the published fluence values for the
four channels of the LAD detector. 
Following
the tradition of radio astronomers (see Lee and Petrosian, 1996), we
divide out the $-3/2$ power law dependence of the number on the
total fluence.  In this way, deviation from
the horizontal line suggests deviation from
a homogeneous, isotropic, static, Euclidean distribution.
In these figures, we have included all 518 bursts available - including
those whose spectral fits gave high values of $\chi^{2}$ - so as
not to underestimate any part of these distributions.
 We have also applied the same procedure
to the subsample of bursts with acceptable values of $\chi^{2}$ and find
that the distributions are similiar to those of the complete
sample within about 30 \%. 
%
Note that the fluence distributions in Figures 4(a) and 4(b) are
unlike the counts of other well known extragalactic
sources such as galaxies, radio sources and AGNs or quasars. The transition
from 3/2 power law is too abrupt and the slope beyond this transition is
nearly constant. Clearly, some extraordinary evolutionary
processes are at work. We explore this in more detail in the
next section.

\subsection{Rate Evolution}
 There have been several detailed analyses 
of the log$N$-log$S$ distribution for the peak
fluxes of GRBs with inconclusive
results.  This is primarily because 
any observed distribution can be fitted
 to an arbitrary luminosity function and evolution even if one
assumes a cosmological model (see e.g. Rutledge et al. 1995,
Reichert and M\'esz\'aros, 1997,
M\'esz\'aros and M\'esz\'aros, 1995, Hakkila et al., 1996).  Additional
uncertainty associated with these results comes from neglect of
the time bias, spectral bias, and uncertainty in the value of the K-correction.
These source of error are absent when dealing with the total
fluence, which requires no correction for time bias (Lee and
Petrosian, 1996),
includes the spectral bias (described above), and requires no K-correction.
Thus, the relation between the total fluence counts and either the
cosmological models or the luminosity function and its evolution is more straight
forward.  Here we concentrate on the rate evolution of GRBs.

  To obtain
some indication of possible evolution,
we assume several representative values for the total radiated energy
${\cal E}$$_{tot}$ and derive the comoving rate density
from the observed differential counts of the fluences as
\begin{equation}
\rho (z) = (1+z)n(F_{tot})(dV/dz)^{-1} (dF_{tot}/dz)
\end{equation}
where $n(F_{tot})$ is the differential distribution of the total
fluence.  Note that in all of our calculations, we have assumed
$H_{o} = 60 \rm km \ s^{-1} Mpc^{-1}$

  Figures 5(a), 5(b) and 5(c) show $\rho(z)$ for three 
representative cosmological models.  In order to
span the range of possible evolutions, we choose
three extreme  models: a flat matter dominated
model ($\Omega_{m} = 1$, $\Omega_{\Lambda} = 0$), a curvature
dominated model ($\Omega_{m} = 0.2$, $\Omega_{\Lambda} = 0$),
and a flat vacuum energy dominated model 
($\Omega_{m} = 0$, $\Omega_{\Lambda} = 1$). Here, $\Omega_{m}$ is the
density parameter (equal to the ratio of matter density to the 
critical density), and $\Omega_{\Lambda} = \Lambda c^{2}/3H_{o}^{2}$ (where
$\Lambda$ is the cosmological constant).
 Within each figure, $\rho(z)$ is plotted
for four standard candle
energies: ${\cal E}$$_{tot} = 10^{51}$, $10^{52}$, $10^{53}$, $10^{54}$ ergs.
Superposed on each figure is the star formation
rate (SFR) from Madau et. al (1997, solid line), this
rate delayed by $2 \times 10^{9}$ years (dashed-dot line), as well
as the SFR convolved with a distribution of time delays 
 $P(t) \propto t^{-1}$ (dotted curve), which may arise in a neutron
 star or black hole merger scenario
(see e.g. Totani, 1997).   Note that the SFR was
calculated using an $\Omega_{m} = 1$, $\Omega_{\Lambda} = 0$ cosmology
(Madau, 1997); however, the shape of this curve as well as the
2 other curves derived from the SFR are fairly insensitive to the
cosmological model (see e.g., Totani, 1997).

Also marked on each figure are the rate densities of GRB 970508 (circle),
GRB 971214 (cross) and GRB 980703 (triangle), given their measured
redshifts, and the energy ${\cal E}$$_{tot}$ required 
to produce a burst at this redshift
(given its total fluence).
Table II lists this required energy.  Again, because we
have only obtained a lower limit to the total fluence of GRB 980703,
we only list a lower limit to its required radiated energy.

\begin{center}
\begin{tabular} {lccccr} \hline \hline
Burst & Redshift &  $F_{tot}$ $ergs/cm^{2}$ & Cosmology & Required ${\cal E}$$_{tot}$ $ergs$ \\ \hline
GRB 970508   & 0.83  & $7.0 \times 10^{-6}$ & $\Omega_{m} = 1 \
\Omega_{\Lambda} = 0$ & $1.0 \times 10^{52}$ \\
GRB 970508   & 0.83  & $7.0 \times 10^{-6}$ & $\Omega_{m} = 0.2 \
\Omega_{\Lambda} = 0$ & $1.4 \times 10^{52}$ \\
GRB 970508   & 0.83  &    $7.0 \times 10^{-6}$ & $\Omega_{m} = 0 \
\Omega_{\Lambda} = 1$ & $2.5 \times 10^{52}$ \\
GRB 971214   & 3.43  &    $1.2 \times 10^{-5}$ & $\Omega_{m} = 1 \
\Omega_{\Lambda} = 0$ & $1.7 \times 10^{53}$ \\
GRB 971214   & 3.43  &    $1.2 \times 10^{-5}$ & $\Omega_{m} = 0.2 \
\Omega_{\Lambda} = 0$ & $4.4 \times 10^{53}$ \\
GRB 971214   & 3.43  &    $1.2 \times 10^{-5}$ & $\Omega_{m} = 0 \
\Omega_{\Lambda} = 1$ & $1.8 \times 10^{54}$ \\
GRB 980703  & 0.97  &    $> 6.1 \times 10^{-5}$ & $\Omega_{m} = 1 \
\Omega_{\Lambda} = 0$ & $>1.1 \times 10^{53}$ \\
GRB 971214   & 0.97  &    $> 6.1 \times 10^{-5}$ & $\Omega_{m} = 0.2 \
\Omega_{\Lambda} = 0$ & $>1,6 \times 10^{53}$ \\
GRB 971214   & 0.97  &    $> 6.1 \times 10^{-5}$ & $\Omega_{m} = 0 \
\Omega_{\Lambda} = 1$ & $>3.2 \times 10^{54}$ \\
 \hline \hline
\end{tabular}
\end{center}

In all cases, the curve corresponding
to ${\cal E}$$_{tot} = 10^{51}$ ergs is ruled out by the observational
data.  The ${\cal E}$$_{tot} \geq 10^{52}$ ergs curve accommodates GRB 970508,
while GRB 971214 and GRB 980703 require ${\cal E}$$_{tot} > 10^{53}$ ergs.
In the 
$\Omega_{\Lambda} = 1$ case, the required ${\cal E}$$_{tot}$ $ > 10^{54}$ ergs
for GRB 971214 begins to exceed
theoretical limits, which may be evidence against
this cosmological model or evidence for strong
beaming of the burst radiation.  
Our results suggest that the total gamma-ray radiated energy
 is not a standard candle and is spread over at least one decade.  We
emphasize that these results depend critically on 
the total fluence, which in turn depends on accurate values
for the spectral parameters for these bursts.  Higher resolution
fits to the data as well as more measurements of the redshifts of
bursts will allow us to constrain the curves more definitely.

Finally, note that none of the three curves for $\rho (z)$ 
follow the star formation
rate curve, as one would expect in many GRB models such as the merger 
or hypernovae models.  The density evolution for GRBS in unlike
that for AGN's or ordinary galaxies as well.  The results agree - at
least qualitatively - with those of Totani (1998).  However, this is
in contradiction with recent claims that the GRB rate follows the
star formation rate (Wijers et al., 1998).

\section{Summary and Conclusion}
 The primary aim of this paper has been to investigate the cosmological
   evolution of GRBs, which involves understanding the distributions
   of the spectral parameters as well as the total fluence of the burst.
 Selection effects limit the information we can observe to explore
 these distributions.  
 
  We have presented new non-parametric
 methods to account for these effects. We have applied these methods
 first to the distribution of the peak of the $\nu F_{\nu}$ spectrum, $E_{p}$.
 We find that this spectral characteristic suffers both upper and lower
 thresholds, although the upper threshold has the dominant effect.
 We correct the observed distribution
  for this truncation and present the parent
 distribution of $E_{p}$.  This contains a large number of
high $E_{p}$ bursts not evident in the observed distribution, which
is in qualitative agreement with the SMM results of Harris and Share (1998b).
This is important for studying correlations between $E_{p}$ and
other burst characteristics, as well as for theoretical considerations
concerning GRB models.

   Using the spectral fits for each burst, we determine a 
 rough measure of the total
 gamma-ray fluence and present the GRB source counts based on this measure. 
 We obtain both the differential and cumulative counts of the total
 and observed fluences. 
 
  We convert these distributions to 
  the comoving rate
 density of GRBs for 3 different cosmological models and 4 different
 values of the total gamma ray radiated energies.
 We find that none
 of the curves follow the star formation rate as predicted by
 various GRB models.  More observations
 on the redshifts of GRBs as well as high resolution spectral
 data (to obtain a precise value for the total fluence) are needed
 to substantiate these results.
 
  As a final note, we summarize some of the caveats in
  our 
analysis.
First, the spectral parameters are determined based on
four channel data and are only  
 rough estimates
of the true spectral properties of the burst.  In particular,
because $\beta$ is chosen from an assumed distribution, the correction to the
$E_{p}$ distribution could be overemphasized.  These spectral
parameters also affect the values of the total fluence (since
$F_{tot} = \int F_{\alpha,\beta,A}(E,E_{p}) dE$),  which in turn affect the
density evolution functions for the bursts. Furthermore,
to derive the density curves,
we have assumed that the bursts total fluence is a standard
candle; although this seems to be the most plausible candidate 
for a standard candle, the observations of GRB 970508, 
971214, and GRB 980703 indicate the contrary.  Note, however, that
 we can use
a standard candle upper and lower limit to at least
constrain the possible 
 evolution of GRBs.
Finally, effects such as beaming were not taken into
account in the density evolution analysis.  For a burst
beamed into a solid angle $\Omega$, the total radiated
energy ${\cal E}$$_{tot}$ will decrease by a factor of $\Omega /4 \pi$ from
the isotropic total radiated energy; hence, the observed dispersion in the total
radiated energy (about an
order of magnitude or higher) could be due to variation in the degree of beaming.
Meanwhile, beaming causes the number of burst
occurances over some time interval to increase by a factor
of $4 \pi / \Omega$.  We will address the issue of how beaming
can affect the cosmological rate evolution in a future publication.

We would like to acknowledge P. Madau for the star formation
rate data used to generate the three SFR curves in 5(a),5(b),
and 5(c).  We would also like to thank B. Efron for help
in implementing the statistical methods discussed in the text. 
Finally, N.M. Lloyd would like to thank members of the BATSE
team for many useful discussions, and their hospitality during
her visit to MSFC.

\appendix{
\section{Appendix}
Here, we describe the results of simulations which
exhibit the technique and demonstrate the accuracy of the
procudure used for
correcting doubly truncated data.
For this analysis, we use the Band spectrum
displayed in equation 5.
We pull $E_{p}$ from a uniform distribution
between 5 and 35 keV.
$\alpha$ is also drawn from a uniform distribution ranging
between -0.9 and 5.9, as was $\beta$ which ranged from
-9.0 to -2.1.
Finally, our normalization parameter $A$ is taken from a uniform
distribution from $12\times 10^{-9}$ to $18 \times 10^{-9} \rm ergs/cm^{2}$.
 Note that we are not trying to simulate the
 real distribution
 of any of these parameters - we choose these
 distributions for the purpose
 of displaying selection effects and to what degree our method
 can account for them.

 Once we have the spectral parameters, we can
 calculate the fluence in each of the four BATSE LAD energy
 channels.  The limiting fluence of each burst
 is chosen from a gaussian distribution with a mean
 of $6 \times 10^{-11} \rm ergs/cm^{2}$ and a standard deviation of
 $6 \times 10^{-10} \rm ergs/cm^{2}$
 (note again, this is not an
 attempt to simulate the actual BATSE detector response).  We
 then ask the question:  How many of the 1000 simulated bursts have
 their fluence in channels 2 and 3, greater than the limiting fluence?
 We find 554 bursts survived this cut - we will call this the observed
 sample.
 We then proceed to apply the method described in \S 2 to our observed
 sample, to see if we could get back the original distribution of $E_{p}$'s.
 Figure 6 shows 
 the the parent sample, the observed sample, and the correction
 to the observed sample.  As evident, the correction accounts
 for the truncated data, and approximately reproduces
 the original sample.  See \S 2 for a full
 description of the technique.}

\newpage

\begin{figure}
\centerline{\psfig{file=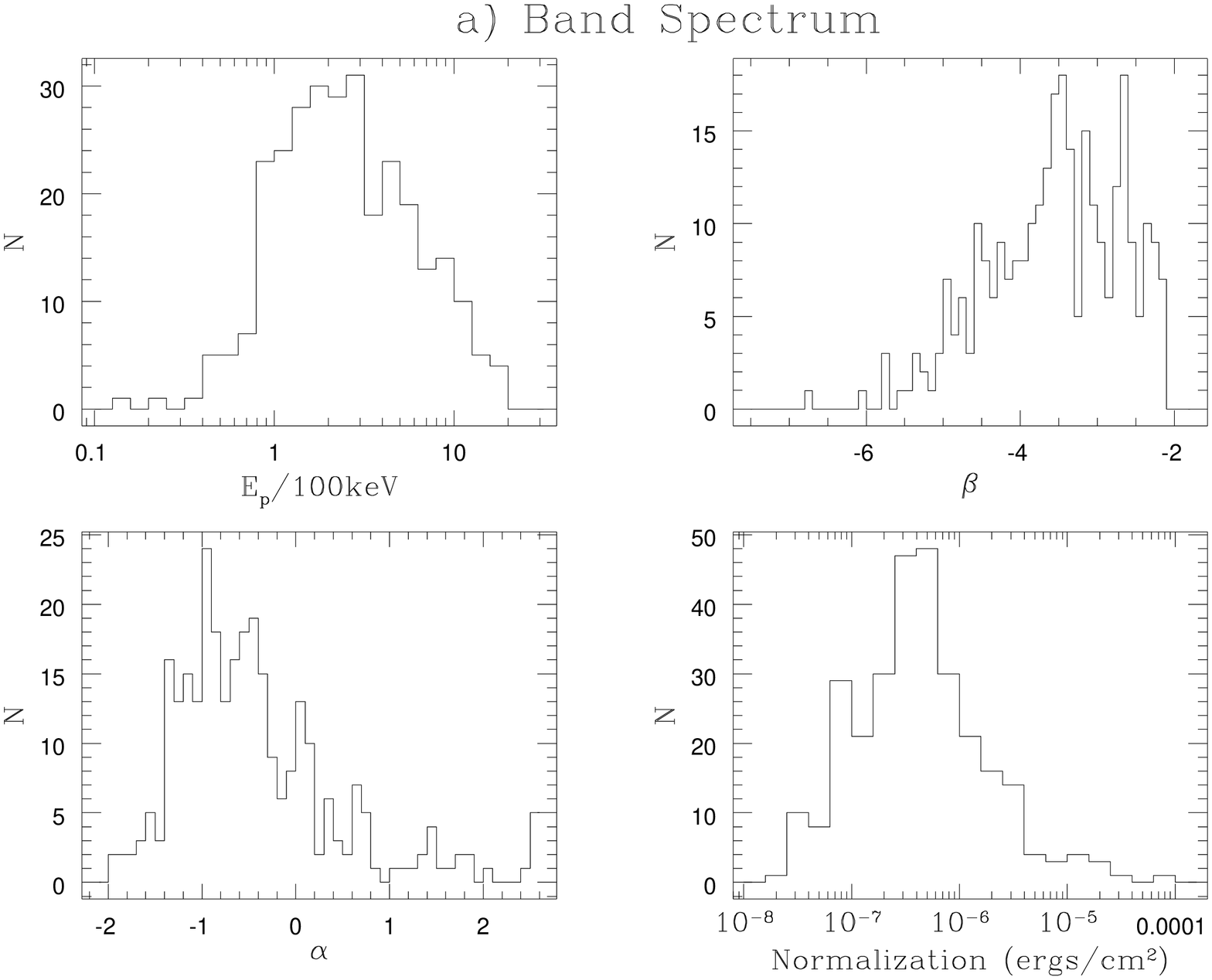,width=6.8cm,height=6cm,angle=270}
\label{(a)}
\psfig{file=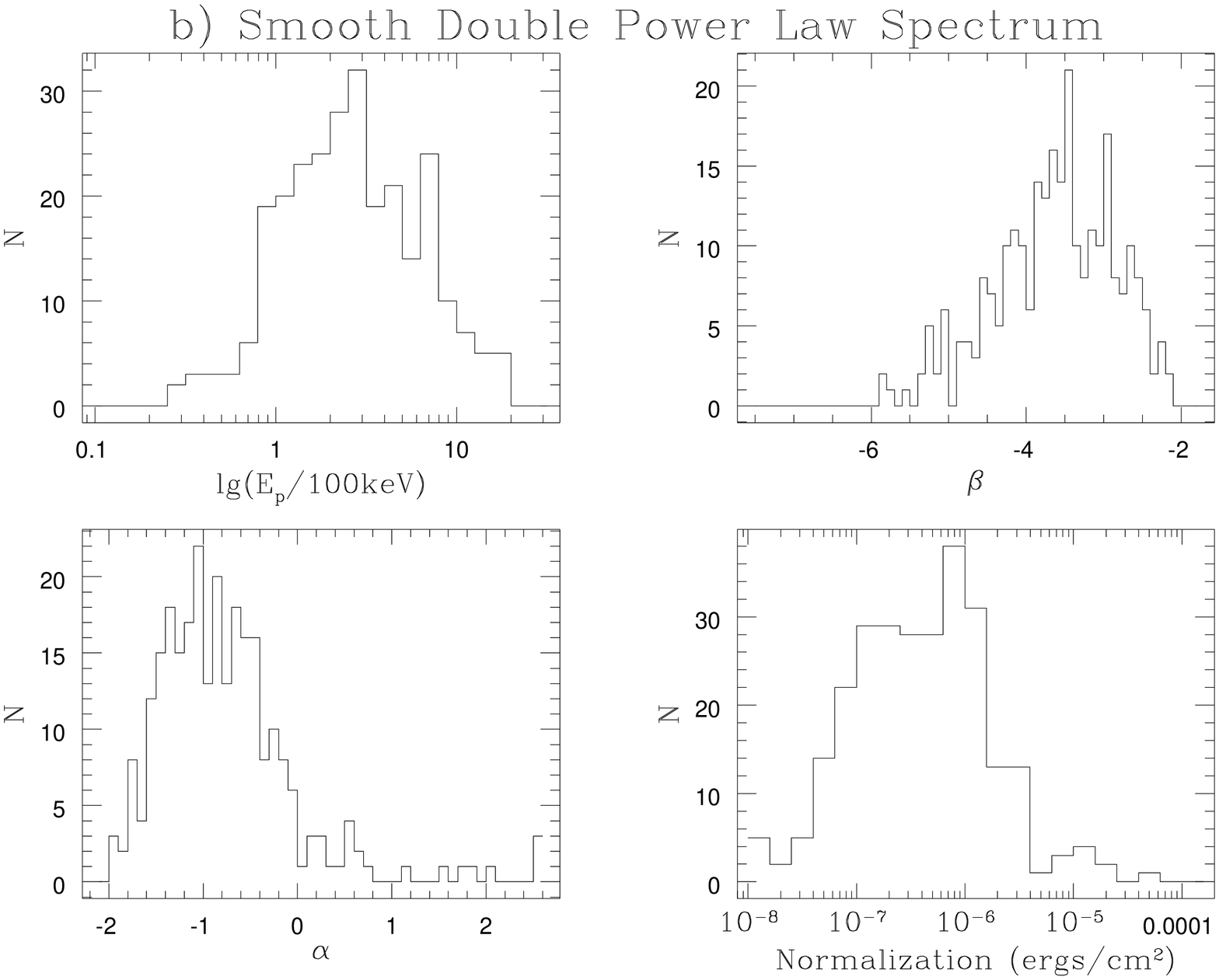,width=6.8cm,height=6cm,angle=270}}
\label{(b)}
\caption{The distribution of the spectral parameters
determined from a) the Band spectrum and
b) a smooth double power-law spectrum respectively. Note
that $\beta$ is drawn from the same truncated gaussian
distribution in both cases, and is not a fitted parameter.
}
\end{figure}

\begin{figure}
\label{2}
\centerline{\psfig{file=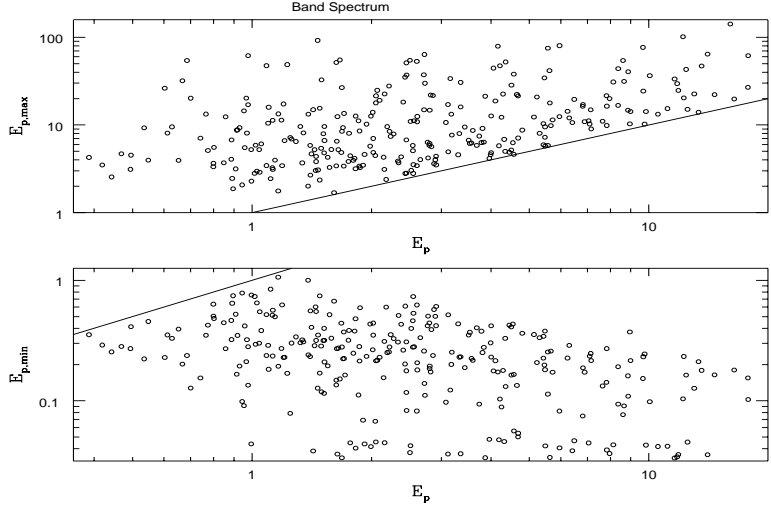,height=7cm,width=11cm,angle=270}}
\caption{
The maximum
and minimum values of $E_{p}$ for the Band spectrum.  Note that the truncation
is more severe from above than below, which
indicates that observations may miss a population of
GRBs with high $E_{p}$.}
\end{figure}

\begin{figure}
\label{3}
\centerline{\psfig{file=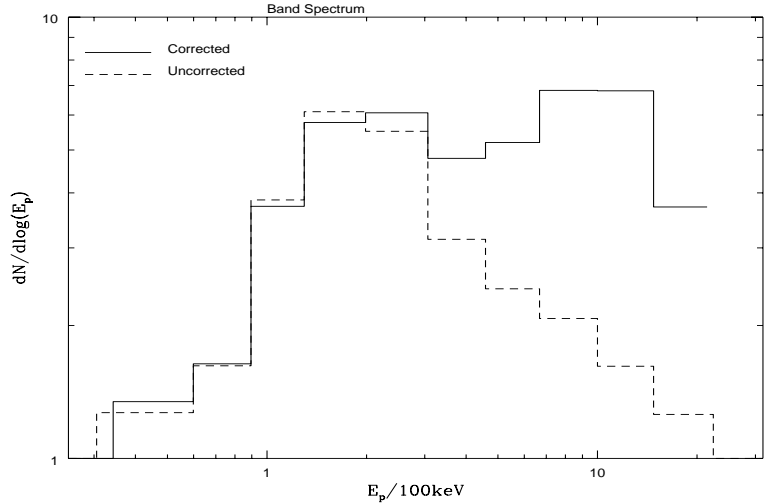,height=7cm,width=11cm,angle=270}}
\caption{ The observed and ``corrected''
distribution
of $E_{p}$ for the Band 
spectrum. 
The figure suggests that there is a significant
sample of high $E_{p}$ bursts undetected by BATSE. 
As mentioned in the text, the methods we used will not produce
a correction beyond the raw observed distribution.}
\end{figure}

\begin{figure}
\centerline{\psfig{file=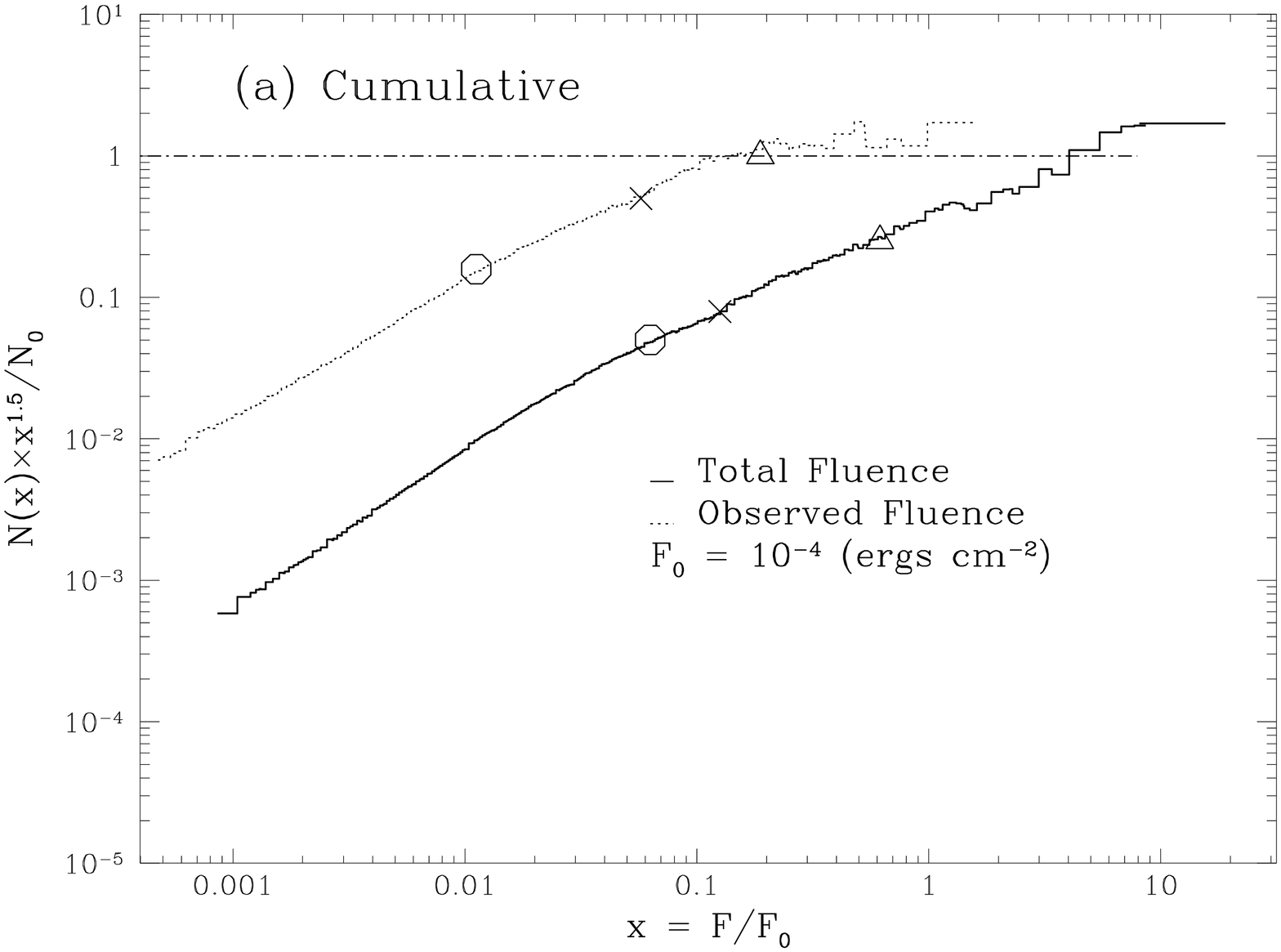,width=6.8cm,height=6cm,angle=270}
\label{(a)}
\psfig{file=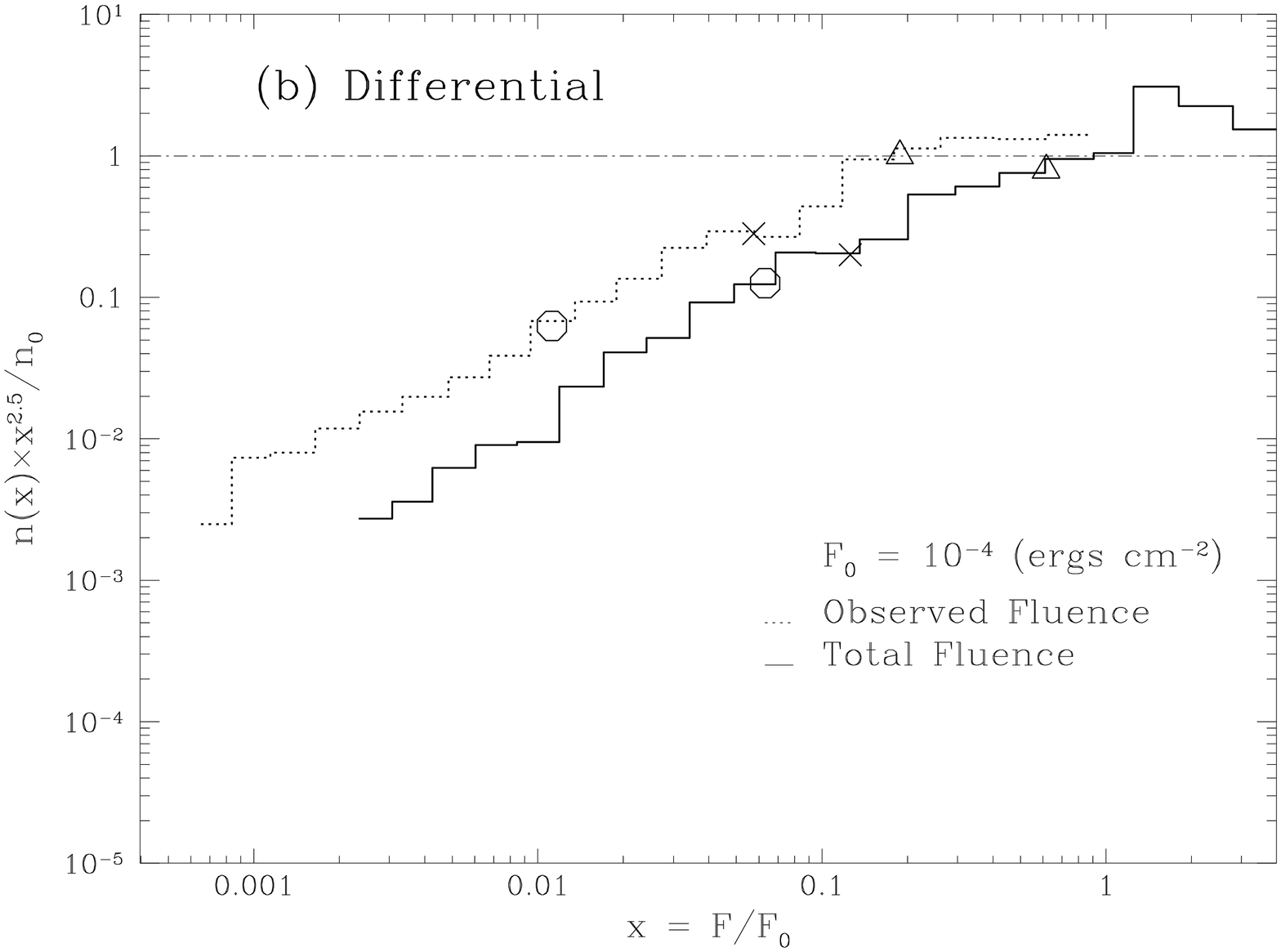,width=6.8cm,height=6cm,angle=270}}
\label{(b)}
\caption{ The cumulative (a) and differential (b)
distribution of both the
observed ($F_{obs}$) and total ($F_{tot}$) fluences.  The circle indicates the
observed and total fluences of GRB 970508, the cross
marks GRB 971214, and the triangle marks GRB 980703.
In each figure, note the similarities between
the two distributions; they both display an abrupt break from the
HISE dependence (see text), indicating presence of strong evolutionary
processes.
}
\end{figure}

\begin{figure}
\label{5(a),5(b),5(c)}
\centerline{\psfig{file=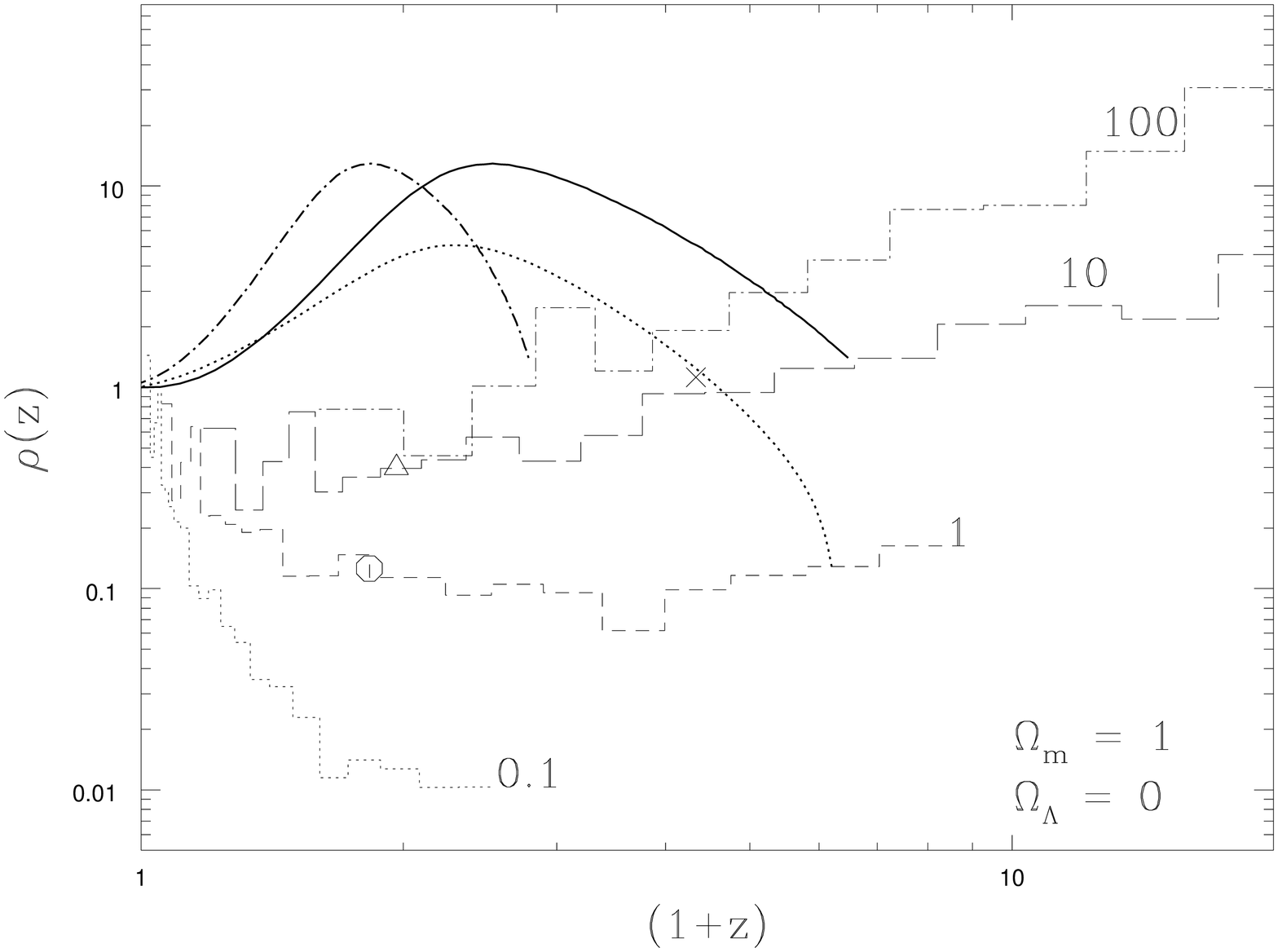,height=5.5cm,width=13cm,angle=270}}
\centerline{\psfig{file=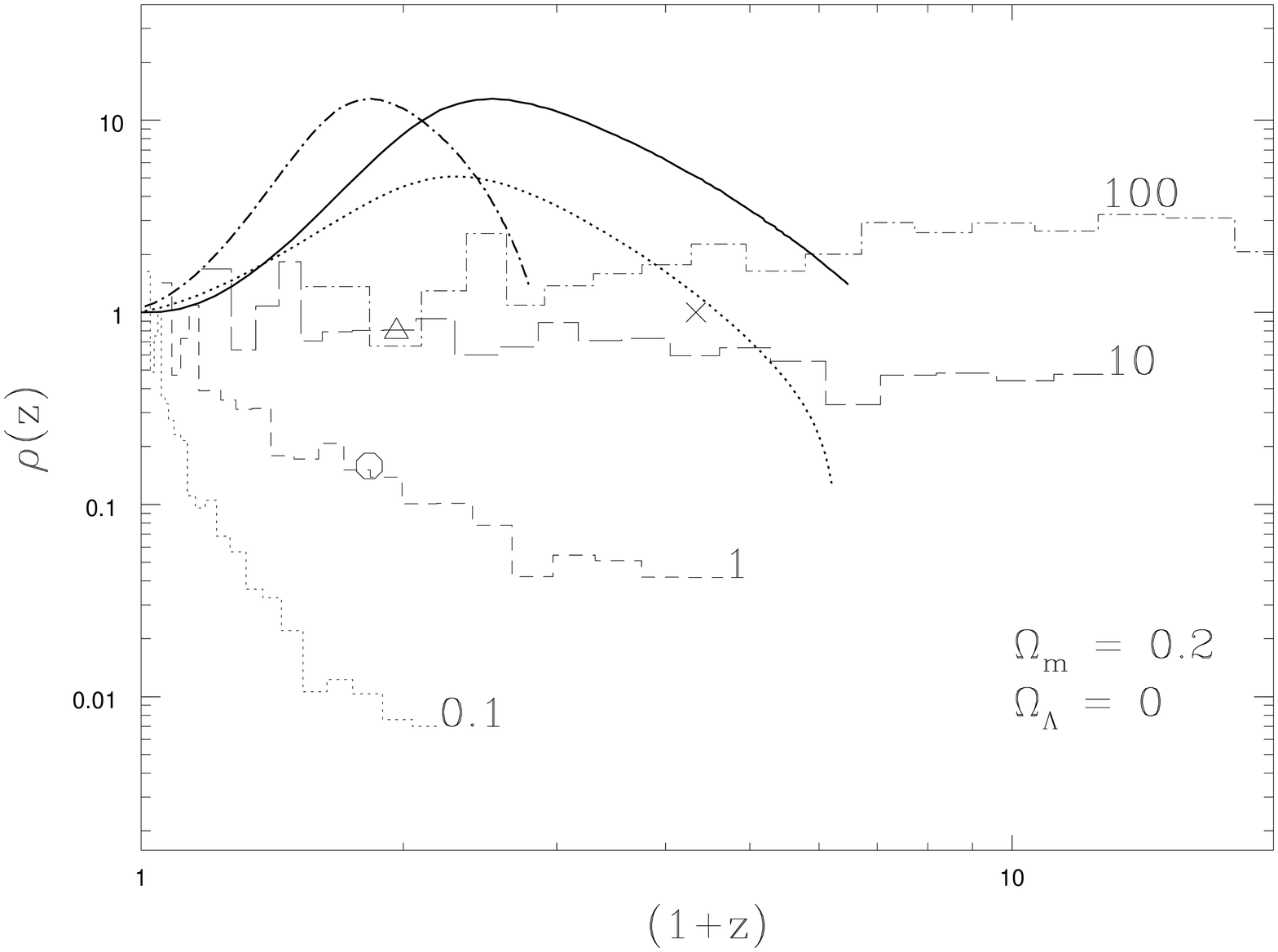,height=5.5cm,width=13cm,angle=270}}
\centerline{\psfig{file=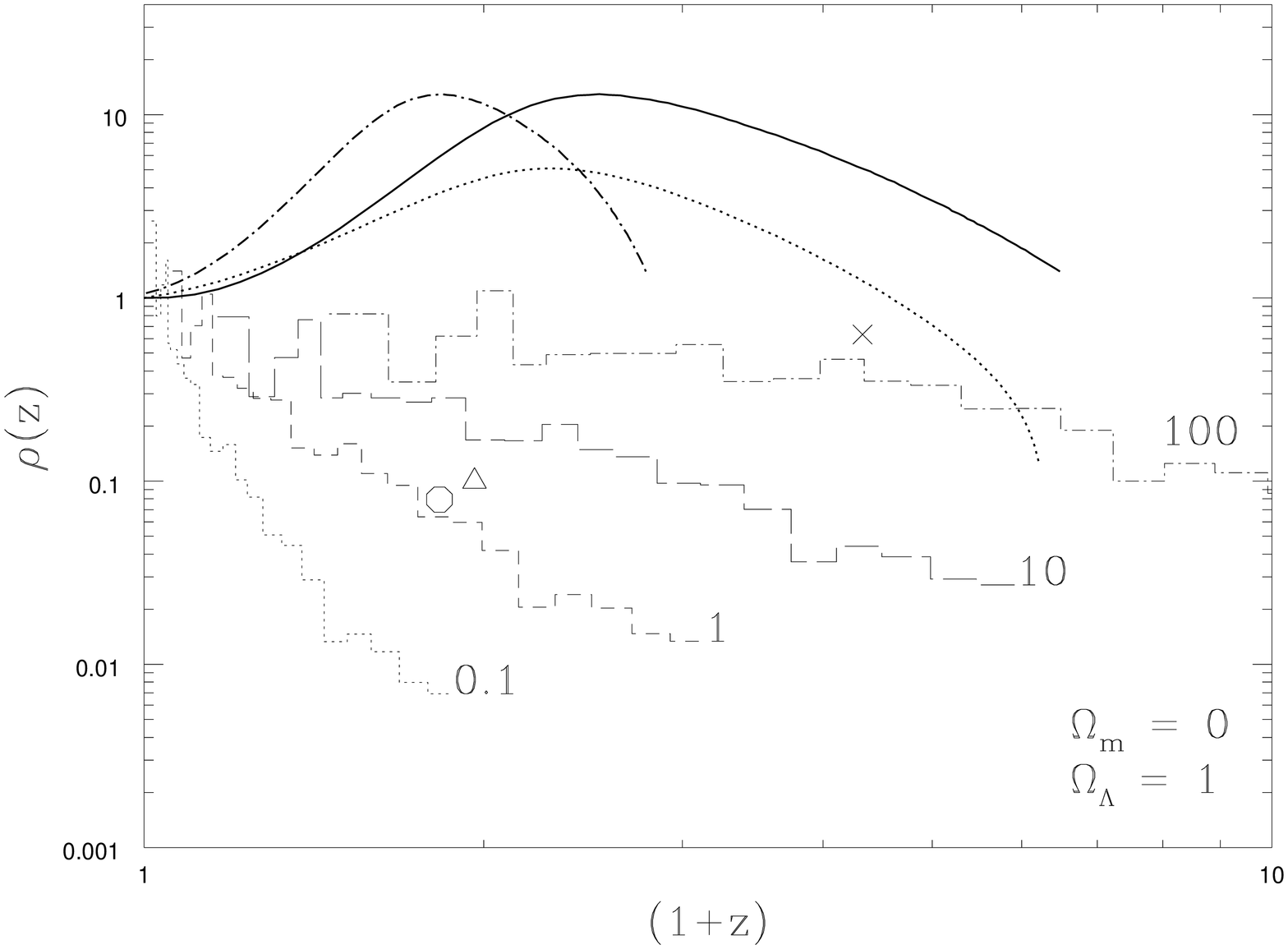,height=5.5cm,width=13cm,angle=270}}
\caption{ The comoving rate density versus redshift $z$
for 3 different cosmological models indicated by the values
of $\Omega_{m}$ and $\Omega_{\Lambda}$.  Within each plot, we show
the rate density for 4 different standard candle energies; the numbers
0.1, 1, 10, and 100 mark the total gamma-ray radiated energy in 
units of $10^{52}$ ergs.
Note that these curves are significantly different
from: 1) the star formation
rate (SFR) from Madau et al., 1997 (solid line), 2) the star formation rate
 with a time
delay of $2 \times 10^{9} $ yr (dot-dash line), and 3) the star formation rate
convolved with a distribution of time delays (dotted line, Totani et al., 1997).
GRB 970508 (circle),
GRB 971214 (cross) and GRB 980703 (triangle) are indicated, given their measured
redshifts and the energy ${\cal E}$$_{tot}$ required 
to produce a burst at this redshift
(given its total fluence, see text).  Note that
no single standard candle energy can accomadate the observed redshifts
of these bursts, in any of the three models.
 }
 \end{figure}

\begin{figure}
\label{6}
\centerline{\psfig{file=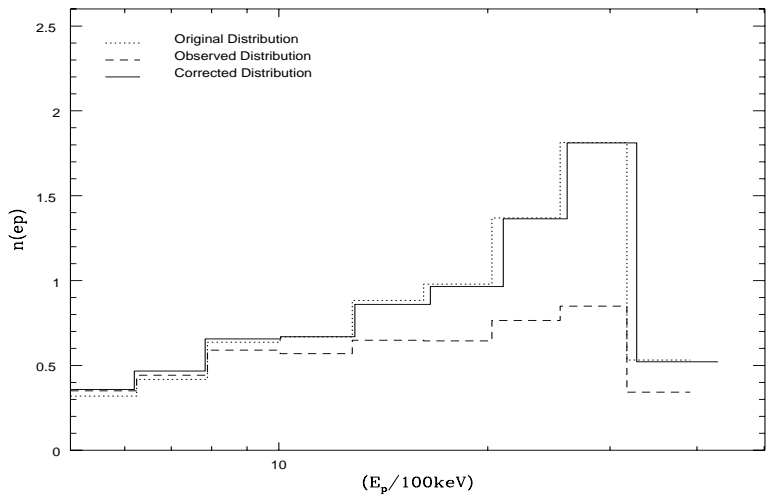,height=7cm,width=11cm,angle=270}}
\caption{ The original $E_{p}$ distribution 
(solid line), the observed $E_{p}$
distribution given the criterion $F_{obs} > F_{lim}$ (dashed line),
and the corrected distribution using the method described
in \S 2 of the text (dotted line).  These distributions are
obtained from the simulations described
in the Appendix.
Note that the procedure recovers the original distribution almost exactly.  
Any differences are primarily due to differences in the binnings of the
distributions.}
\end{figure}

\end{document}